\documentclass[
groupedaddress,
twocolumn, 
notitlepage,
aps, pra, 10pt]{revtex4-1}

\usepackage{graphicx}
\usepackage{physics}
\usepackage{amsmath}
\usepackage{amssymb}
\usepackage{cancel}
\usepackage{empheq}
\usepackage{xcolor}
\usepackage[colorlinks,citecolor=blue,urlcolor=blue,linkcolor=blue,hypertexnames=true]{hyperref}
\usepackage{tabularx}
\newcolumntype{A}{>{\centering\arraybackslash}X}
\usepackage{array, makecell}

\newcommand{\etal}{\emph{et al. }}

\begin{document}

\title{\textbf{Energy and photon centroids of spatiotemporal light pulses \\ and consequences for their
intrinsic orbital angular momentum}}
\author{Titouan Gadeyne}
\author{Thierry Ruchon}
\affiliation{%
Universit\'e Paris-Saclay, CEA, LIDYL, 91191 Gif-sur-Yvette, France
}

\begin{abstract}
An ongoing debate surrounding the intrinsic transverse orbital angular momentum (OAM) attributed to spatiotemporal optical vortex (STOV) light pulses has raised the energy and photon-density centroids as two alternative frameworks to define the center of a free electromagnetic wavepacket. 
We herein derive, within a single formalism, lowest-order expressions for the positions and velocities of both centroids, directly applicable to an arbitrary scalar, near-paraxial and quasi-monochromatic pulse envelope formulated in space-time. Examining pulses with structures including temporal chirp, wavefront rotation, Gaussian-type STOVs and tilted lobulated profiles, we illustrate general principles underlying the shifts and propagation dynamics of the two centroids, and discuss how these properties underpin the value and conservation of intrinsic OAM referenced to either centroid. 
Finally, despite widespread use of the notion of intrinsic OAM \emph{per photon}, we argue that in neither framework could individual STOV photons be said to carry a well-defined quantum of intrinsic transverse OAM, as such quantities do not write as expectation values of a single operator for which STOV states could form a basis of orthogonal eigenmodes. This contrasts with the well-established quantization scheme for the total longitudinal OAM of light in spatial vortex beams, bearing implications for the interpretation of future experiments involving STOV pulses in quantum optics, nonlinear wave-mixing or light-matter interactions.
\end{abstract}

\date{\today}

\maketitle

\section{Introduction}
\label{sec-introduction}

The orbital angular momentum (OAM) contained in structured electromagnetic fields continues to puzzle optical scientists, decades after the seminal discovery by Allen \etal that so-called vortex beams of light possess OAM oriented parallel to their direction of propagation \cite{allen_1992_orbital, andrews_2012_angular, shen_2019_optical}. 
As a prime example, spatiotemporal optical vortex (STOV) light pulses have recently garnered considerable attention \cite{
hancock_2019_free,
chong_2020_generation,
huo_2024_observation,
chen_2023_spatiotemporal,
wan_2023_optical}. These wavepackets exhibit a phase vortex embedded in the space-time domain,
endowing them with OAM oriented transverse to their axis of propagation (t-OAM). However, evaluating the t-OAM transported by light pulses structured in space-time has sparked a wealth of subtle debates \cite{
bliokh_2012_spatiotemporal,
bliokh_2021_spatiotemporal, 
hancock_2021_mode,
bliokh_2023_orbital, 
porras_2023_transverse, 
hancock_2024_spatiotemporal,
porras_2024_comment,
porras_2024_commenta,
porras_2024_clarification}. 
One difficulty lies in the fact that the OAM $\vb{L}$ is defined with respect to the frame origin: displacing this origin as $\vb{r} \to \vb{r} - \vb{r}_0$ changes the OAM as $\vb{L} \to \vb{L} - \vb{r}_0 \cross \vb{P}$, where $\vb{P}$ is the total linear momentum of the field \cite{bliokh_2015_transverse}. While this leaves the component of $\vb{L}$ parallel to $\vb{P}$ unchanged, its transverse part can be considerably affected, even by subwavelength shifts of the origin. This motivates the splitting
of optical OAM into \emph{extrinsic} (whose value does depend on the choice of origin) and \emph{intrinsic} (whose value does not) parts, as
$\vb{L}=\vb{L}_e + \vb{L}_i$.
\\
Broadly speaking, the extrinsic OAM $\vb{L}_e$ of an optical pulse is due to the lever arm the propagation of its center creates with respect to the origin, which can be non-zero even for simple (\textit{e.g.} plain Gaussian) wavepackets. In contrast, an intrinsic part $\vb{L}_i$ only arises for wavepackets endowed with a more complex internal structure, inside which electromagnetic energy flows with a sense of rotation around a center point. Unfortunately, many procedures could be imagined to construct an origin-independent quantity from the total OAM $\vb{L}$ \cite{bliokh_2015_transverse}. In recent years, two main strategies have been put forward, which we now briefly review. 
\\

In a first approach (explored rigorously by Porras in \cite{porras_2024_clarification}), the splitting is achieved by referencing OAM to the \emph{energy centroid} $\vb{r}_\text{EC}$ of the pulse, as
\begin{align}
    \vb{L}_{i,\text{EC}}
    =
    \vb{L} - \vb{r}_\text{EC} \cross \vb{P},
\label{eq:Li_EC}
\end{align}
where $\vb{r}_\text{EC}$ is the barycenter of the electromagnetic energy density \cite{zangwill_2012_modern}.
As a displacement of the origin by $\vb{r}_0$ also shifts the centroid as $\vb{r}_\text{EC} \to \vb{r}_\text{EC} - \vb{r}_0$, this quantity is indeed origin-independent. With this definition, STOV pulses with circular symmetry in space-time and topological charge $\ell$ were found to possess \textit{half-integer} average intrinsic t-OAM per photon, $\hbar \ell/2$ \cite{porras_2024_clarification}. 
Since $\vb{P}$ and $\vb{L}$ are conserved in time, and the energy centroid always moves at a constant velocity $\vb{v}_\text{EC} = \partial_t \vb{r}_\text{EC} \propto \vb{P}$ \cite{zangwill_2012_modern}, the quantity $\vb{L}_{i,\text{EC}}$ is itself conserved. In this respect, it was successfully applied by Hancock \etal to describe experiments involving spatiotemporally perturbed pulses, detected after some propagation \cite{hancock_2024_spatiotemporal, hancock_2025_transverse}.
\\
An alternative strategy proposed by Bliokh \cite{bliokh_2023_orbital} references $\vb{L}$ to the so-called \emph{photon centroid} $\vb{r}_\text{PC}$,
\begin{align}
    \vb{L}_{i,\text{PC}}
    =
    \vb{L} - \vb{r}_\text{PC} \cross \vb{P}. 
\label{eq:Li_PC}
\end{align}
where $\vb{r}_\text{PC}$ is obtained as the expectation value of the position operator against a \emph{photon wavefunction} \cite{good_1957_particle,
bialynickibirula_1996_photon, smirnova_2018_relativistic, bliokh_2023_orbital}. It is an optical analogue to the ``particle-number" centroid of a material body, which has already been used alongside the energy centroid to gain insight into the description of angular momentum phenomena \cite{bliokh_2012_relativistic}.
In this picture, STOV pulses with circular symmetry have \textit{integer} intrinsic t-OAM per photon $\hbar \ell$, and the dependence of t-OAM on the space-time eccentricity of the pulse acquires an elegantly general form, coinciding with that for the longitudinal OAM of regular vortex beams \cite{bliokh_2023_orbital}. 
For an arbitrary wavepacket the photon centroid may not travel in the direction of $\vb{P}$ \cite{hancock_2024_spatiotemporal, bliokh_2025_ehrenfest}, hence this splitting of $\vb{L}$ is generally not time-independent.
If the conserved nature of $\vb{L}_{i,\text{EC}}$ makes it more convenient for describing free-space propagation, little is known about the possible relevance of either framework as a tool to interpret \textit{e.g.} interactions of STOVs with matter. Up to this point, the t-OAM of STOVs has only been detected indirectly, by computing it from a measurement of the light field itself: in this case, both formalisms \eqref{eq:Li_EC} and \eqref{eq:Li_PC} could equivalently be applied, yielding very different yet self-consistent pictures. 

Since these proposals, attempts at comparing energy and photon centroids have remained difficult, partly due to the distinct mathematical formalisms they involve: while $\vb{r}_\text{EC}$ is commonly defined as an integral over space $(x,y,z)$ \cite{zangwill_2012_modern}, expressing $\vb{r}_\text{PC}$ requires a reciprocal-space formulation (\emph{i.e.} plane-wave expansion) of the fields -- for this reason, photon-centroid quantities have been obtained for theoretical fields with simple wavevector spectra, such as Bessel pulses \cite{
smirnova_2018_relativistic,
bliokh_2023_orbital, bliokh_2023_photon}. Both centroids can prove cumbersome to compute however, as structured pulses are commonly prescribed with a complex amplitude envelope known in space-time $(x,y,t)$ at a single coordinate $z$ along their axis of propagation. Such a space-time formulation is arguably more in line with experimental manipulation of ultrashort pulses \cite{porras_2024_clarification}, which are typically shaped in the frequency domain (thus controlling their \emph{temporal} structure more directly than their \emph{longitudinal} one), and probed with detectors set at a fixed $z$ position. \\

In this work, we give a rigorous procedure to express electromagnetic quantities defined in reciprocal space directly in terms of a near-paraxial, quasi-monochromatic and linearly polarized pulse envelope specified in space-time. We obtain, within a single formalism, formulae for the two centroids, their velocities and associated intrinsic OAM -- ample details on our calculations are provided in the Supplementary Material (SM) \cite{supplementary}. These general expressions allow us to illustrate centroid trajectories and their consequences for intrinsic OAM on a variety of pulses with structures including temporal chirp, wavefront rotation, Gaussian-type STOVs and tilted lobulated profiles. 
Lastly, we reevaluate the notion of intrinsic OAM \emph{per photon}, observing that these constructions -- regardless of the chosen centroid -- are not quantities that individual photons in spatial or spatiotemporal optical vortices could carry in well-defined quanta: intrinsic OAM (transverse or longitudinal) is not generated by a single operator for which vortex states could be the eigenmodes. This differs from how the total longitudinal OAM of light is quantized in spatial vortex beams, which has offered an
unprecedented platform for quantum experiments \cite{mair_2001_entanglement,
molinaterriza_2007_twisted,
erhard_2017_twisted}. From the perspective of quantum optics, STOV pulses should not be expected to function as analogues of regular vortex beams.

\section{Theoretical framework}
\label{sec-theory}

\subsection{Reciprocal-space formulation of relevant electromagnetic quantities}
\label{subsec-relevant}

We consider classical electromagnetic wavepackets in free space with finite energy, for which fields vanish sufficiently fast at infinity. The real electric field $\boldsymbol{\mathcal{E}}(\vb{r},t)$ may be expressed as a sum of homogeneous plane waves \cite{zangwill_2012_modern}
\begin{align}
    \boldsymbol{\mathcal{E}}
    & =
    \Re{\vb{E}}
    ,
    \;\;\;
    \vb{E}(\vb{r}, t)
    =
    \int \frac{\dd \vb{k}}{(2\pi)^{3/2}} ~
    \widetilde{\vb{E}}(\vb{k}) e^{i \vb{k} \vdot \vb{r}-i \omega(\vb{k}) t}
    \label{eq:CAS}
\end{align}
with wavevectors $\vb{k}$, frequencies $\omega(\vb{k}) = c\abs{\vb{k}}$, and complex amplitudes $\widetilde{\vb{E}}(\vb{k})$ verifying transversality,
\begin{align}
    \vb{k} \vdot \widetilde{\vb{E}}(\vb{k}) = 0.
    \label{eq:transversality}
\end{align}
The complex field $\vb{E}$ is the analytic-signal representation of $\boldsymbol{\mathcal{E}}$, containing only positive-frequency waves. Using this reciprocal-space formulation of the fields, the total energy $U$, momentum $\vb{P}$ and orbital angular momentum $\vb{L}$ of the wavepacket write \cite{cohen-tannoudji_1953_photons,
li_2009_spin}
\begin{align}
    U 
    & = \frac{\varepsilon_0}{2} \int \dd \vb{k}~
    |\widetilde{\vb{E}}(\vb{k})|^2
    \label{eq:U}
    \\
    \vb{P} 
    & =
    \frac{\varepsilon_0}{2} \int \dd \vb{k}~
    \omega^{-1} \vb{k} |\widetilde{\vb{E}}(\vb{k})|^2
    \label{eq:P}
    \\
    \vb{L}
    & =
    \frac{\varepsilon_0}{2}
    \int \dd \vb{k}~ \omega^{-1}
    \widetilde{\vb{E}}^\star(\vb{k}) 
    \vdot
    (-i \vb{k} \cross \grad_{\vb{k}})
    \widetilde{\vb{E}}(\vb{k}) 
    \label{eq:L}
\end{align}
where \eqref{eq:L} uses the customary notation $[\vb{A} \vdot (\vb{B} \cross \grad) \vb{C}]_i = \sum_{jkm} A_m \epsilon_{ijk} B_j \partial_k C_m$, and $\vb{L}$ is the OAM with respect to the frame origin.
The energy centroid $\vb{r}_\text{EC}$ of the wavepacket (and its velocity $\vb{v}_\text{EC}$) write \cite{bliokh_2023_photon}
\begin{align}
\begin{split}
    \vb{r}_\text{EC}(t) 
    & = 
    \frac{\varepsilon_0}{2}
    \frac{1}{U}
    \int \dd \vb{k}~
    \widetilde{\vb{E}}^\star(\vb{k}) e^{i \omega t}
    \vdot
    (i \grad_{\vb{k}})
    \widetilde{\vb{E}}(\vb{k}) e^{-i \omega t}
    \\
    & = \vb{r}_\text{EC}(t=0) + t \,\vb{v}_\text{EC}
    \\
    \vb{v}_\text{EC}
    & =
    \frac{\varepsilon_0}{2}
    \frac{1}{U}
    \int \dd \vb{k}~
    \frac{c \vb{k}}{\abs{\vb{k}}}
    |\widetilde{\vb{E}}(\vb{k})|^2
    =
    \frac{c^2 \vb{P}}{U}
    \label{eq:rEC_vEC}
\end{split}
\end{align}
where \eqref{eq:rEC_vEC} uses the notation $[\vb{A} \vdot (\grad) \vb{B}]_i = \sum_j A_j \partial_i B_j$. Introducing $(\boldsymbol\Psi |\bullet| \boldsymbol\Psi) = (\varepsilon_0/2)\int \dd \vb{k} ~ \boldsymbol\Psi^\star \vdot (\bullet)\boldsymbol\Psi$, all quantities above rewrite as expectation values with the \emph{photon wavefunction} $\boldsymbol\Psi(\vb{k}) = \widetilde{\vb{E}}(\vb{k}) e^{-i \omega(\vb{k}) t}/\sqrt{\omega(\vb{k})}$ \cite{bialynickibirula_1996_photon, 
good_1957_particle, bliokh_2023_orbital} -- for a given quantity, the contribution from each plane wave is proportional to its density of photons, $\propto |\widetilde{\vb{E}}(\vb{k})|^2/\omega(\vb{k})$. The energy centroid
$\vb{r}_\text{EC} = (\boldsymbol\Psi |\omega i \grad_{\vb{k}}| \boldsymbol\Psi)/(\boldsymbol\Psi |\omega | \boldsymbol\Psi)$ then involves the expectation value of the position operator ($i \grad_{\vb{k}}$ in momentum representation) weighted by the photon energy. This framework allows to naturally define the number of photons $N = (\boldsymbol\Psi | \boldsymbol\Psi)$ (in units of $\hbar$) and the photon-density centroid $\vb{r}_\text{PC} = (\boldsymbol\Psi |i \grad_{\vb{k}} | \boldsymbol\Psi) / (\boldsymbol\Psi | \boldsymbol\Psi)$ \cite{smirnova_2018_relativistic,
bliokh_2023_orbital, bliokh_2023_photon} of the wavepacket, that is explicitly
\begin{align}
    N 
    & = 
    \frac{\varepsilon_0}{2} \int \dd \vb{k}~
    \omega^{-1} |\widetilde{\vb{E}}(\vb{k})|^2
    \label{eq:N}
    \\
    \vb{r}_\text{PC}(t) & = 
    \frac{\varepsilon_0}{2}
    \frac{1}{N}
    \Re
    \int \dd \vb{k}~ \omega^{-1}
    \widetilde{\vb{E}}^\star(\vb{k}) e^{i \omega t}
    \vdot
    (i \grad_{\vb{k}})
    \widetilde{\vb{E}}(\vb{k}) e^{-i \omega t}
    \nonumber
    \\
    & = \vb{r}_\text{PC}(t=0) + t \,\vb{v}_\text{PC}
    \label{eq:rPC_vPC}
    \\
    \vb{v}_\text{PC}
    & =
    \frac{\varepsilon_0}{2}
    \frac{1}{N}
    \int \dd \vb{k}~
    \omega^{-1}
    \frac{c \vb{k}}{\abs{\vb{k}}}
    |\widetilde{\vb{E}}(\vb{k})|^2 
    \nonumber
\end{align}
and the velocity $\vb{v}_\text{PC}$ is generally not aligned with $\vb{P}$ \cite{bliokh_2023_photon}. Evaluation of $\vb{r}_\text{PC}$ from the integral in \eqref{eq:rPC_vPC} requires discarding an imaginary part, although it is not explicit in \cite{bliokh_2023_orbital, bliokh_2023_photon} -- this minor point is briefly discussed in the SM \cite{supplementary}. \\
Finally, average quantities \emph{per photon} are obtained (in units of $\hbar$) upon dividing them by $N$. We note that the wording is sometimes used to mean \emph{per unit energy} \cite{porras_2023_transverse} -- though strictly speaking the two notions are different.
\\

All quantities introduced in this section are real-valued, and involve integrands quadratic in the field components -- considering \eqref{eq:transversality}, all $z$ components can be expressed from the transverse $\perp~= (x,y)$ ones as
$
    \widetilde{E}_z(\vb{k}) 
    =
    -  (\vb{k}_\perp/k_z) \vdot 
    \widetilde{\vb{E}}_\perp(\vb{k})
$,
such that all integrals of interest can be cast in the general form
\begin{align}  
\frac{\varepsilon_0}{2}
\Re
\int \dd \vb{k} ~
    \widetilde{\vb{E}}_\perp^\star(\vb{k}) 
    \vdot
    \mathbb{A}
    \widetilde{\vb{E}}_\perp(\vb{k})
\label{eq:generalA}
\end{align}
where $\mathbb{A}$ is a $2\cross2$ matrix, dependent on $\vb{k}$ and possibly containing momentum-space derivatives acting on $\widetilde{\vb{E}}_\perp(\vb{k})$ \cite{supplementary}.

\subsection{Prescription of a quasi-monochromatic and near-paraxial pulse in space-time}

We now concern ourselves with quasi-monochromatic and near-paraxial pulses, propagating in the positive $z$ direction, and therefore take the fields to admit no backward-propagating waves, that is $\widetilde{\vb{E}}(\vb{k}_\perp, k_z<0) = 0$. Components of the electric field transverse to the $z$-axis are then assumed to be known in space-time at the plane $z=0$, and prescribed under the form 
\begin{align}
    \boldsymbol{\mathcal{E}}_\perp(\vb{r}_\perp, z=0, t)
    = \Re{
    \boldsymbol\psi(\vb{r}_\perp, t) e^{- i \omega_0 t}
    }
\label{eq:prescription}
\end{align}
where $\omega_0 = c k_0$ is a carrier optical frequency and $\boldsymbol\psi = (\psi_x, \psi_y)$ contains complex amplitude envelopes for each transverse component. 
Under quasi-monochromatic conditions, the envelope $\boldsymbol\psi$ is sufficiently narrow-band for negative frequencies to make negligible contributions, and transverse components of $\widetilde{\vb{E}}(\vb{k})$ can be directly obtained from a Fourier transform (FT) of the prescribed complex field as \cite{supplementary}
\begin{align}
    \widetilde{\vb{E}}_\perp(\vb{k})
    \approx
    \frac{c^2 k_z}{\omega(\vb{k})}
    \int \frac{\dd \vb{r}_\perp  \dd \tau}{(2\pi)^{3/2}} ~
    \boldsymbol\psi(\vb{r}_\perp, \tau)
    e^{-i\vb{k}_\perp \vdot \vb{r}_\perp
    +
    i[\omega(\vb{k})-\omega_0] \tau}
\label{eq:PW_amplitude}
\end{align}
After evaluating the possible momentum-space derivatives \cite{supplementary}, all integrals of the type \eqref{eq:generalA} acquire the general FT form
\begin{align*}
\begin{split}
\frac{\varepsilon_0 c}{2}
&
\Re
\int \dd \vb{r}_\perp \dd \tau
\int \dd \vb{r}'_\perp \dd \tau'
\int
\frac{\dd \vb{k}_\perp \dd \omega}{(2\pi)^3}~ 
\\
& e^{i \vb{k_\perp} \vdot [\vb{r}'_\perp - \vb{r}_\perp] }
e^{-i [\omega - \omega_0] [\tau'-\tau] }
\boldsymbol\psi^\star(\vb{r}'_\perp, \tau')
\vdot
\mathbb{B}
\boldsymbol\psi(\vb{r}_\perp, \tau)
\end{split}
\end{align*}
where $\mathbb{B}$ is a $2\cross2$ matrix that may involve $\vb{k}_\perp$, $\omega$, $\vb{r}_\perp$ and $\tau$, but not the primed variables.\\
The small parameters pertaining to our paraxial and quasi-monochromatic approximations are
\begin{align*}
    \kappa_x = \frac{k_x}{k_0},
    \;\;\;
    \kappa_y = \frac{k_y}{k_0},
    \;\;\;
    \kappa_\omega = \frac{\omega-\omega_0}{\omega_0}.
\end{align*}
For wavepackets having similar extents along the transverse and longitudinal directions (such as STOV pulses with circular symmetry in space-time) those three parameters have a comparable magnitude, denoted $\kappa \ll 1$. Components of $\mathbb{B}$ are then expanded as a power series in $\kappa$ up to the desired order of approximation, whereafter integration over $\vb{k}_\perp$, $\omega$, $\vb{r}'_\perp$ and $\tau'$ can be systematically performed using standard properties of the FT and derivatives of the Dirac delta function \cite{supplementary}, in similar fashion to the calculations of \cite{li_2009_spin,
mansuripur_2011_spin}. 
\\

For the rest of this work, we restrict ourselves to pulses with homogeneous linear polarization along the $x$-axis, $\boldsymbol\psi \to (\psi, 0)$, thus working with a single scalar envelope.
To write the final expressions concisely, we introduce the following braket notation
\begin{align*}
    \expval{\bullet}{\psi} 
    = 
    \frac{\varepsilon_0 c}{2} \int \dd \vb{r_\perp} \dd \tau ~
    \Re{\psi^\star(\vb{r}_\perp,\tau) \bullet \psi(\vb{r}_\perp,\tau)}.
\end{align*} 
All quantities defined in reciprocal space in section \ref{subsec-relevant} can therefore be expressed from the expectation values of certain operators acting on the space-time envelope $\psi$ of the light pulse. 
Finally, we shall write the corresponding normalized expectation values as
\begin{align*}
    \expval{\bullet}
    = 
    \frac{\expval{\bullet}{\psi}}{\braket{\psi}}.
\end{align*}

\section{Results and discussion}

We now present formulae expressing all quantities of section \ref{subsec-relevant} directly from the pulse envelope $\psi$ - details on their derivations can be found in the SM \cite{supplementary}. We pursue calculations up to first order in $\kappa$ except when indicated otherwise -- this level of approximation matches that used in previous studies on the t-OAM of STOVs.  
\\
We find, for the energy and photon number
\begin{align}
    U & \approx 
    \braket{\psi}
    \label{eq:U_1st_order}
    \\
    N\omega_0 & \approx 
    U
    - 
    \frac{1}{k_0}
    \expval{i\partial_\tau/c}{
    \psi}
    \label{eq:N_1st_order}
\end{align}
At this level of approximation, $U$ only captures the energy in transverse field components, as the square of the longitudinal field is of order $\kappa^2$. In a polychromatic field, the photon number generally differs from $U/\omega_0$, since it depends on the specific shape of its frequency spectrum -- 
higher-order terms in \eqref{eq:N_1st_order} would involve higher-order moments $\expval{(i\partial_\tau)^n}{
    \psi}$ of the spectral energy density.
\\
The linear momentum reads
\begin{align}
    \vb{P} & \approx 
    \frac{1}{c} 
    \left[ U 
    \hat{\vb{z}}
    +
    \frac{1}{k_0}
    \expval{
    \begin{pmatrix}
        -i\grad_\perp \\
        0 
    \end{pmatrix}
    }{\psi}
    \right]
    \label{eq:P_1st_order}
\end{align}
with the upper component of a column vector being understood to contain the two Cartesian components transverse to $z$, while the lower component concerns the temporal dimension. Here the first-order correction term captures pulse propagation at a small angle with respect to the $z$-axis. 
\begin{figure}[b!]
    \centering
    \includegraphics[width=\linewidth]{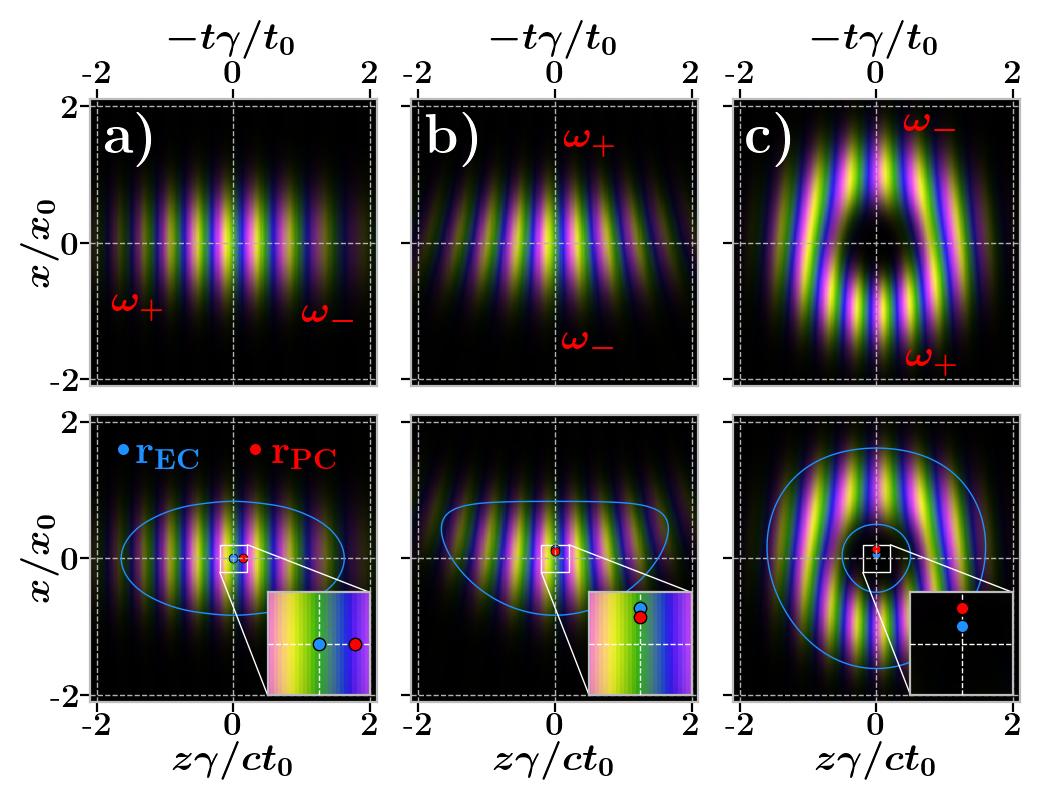}
    \caption{
    \textbf{Shifts of the energy and photon centroids.} \\
    Top: space-time pulse $\psi(x,t)e^{-i\omega_0 t}$ realized at the plane $z=0$. Throughout this article, complex fields are rendered with brightness proportional to their squared modulus and phase mapped to a cyclic color wheel. $\omega_+$ (resp $\omega_-$) indicate the regions of higher (resp. lower) local frequency. \\
    Bottom: complex electric field component $E_x(x,z)$ at $t=0$, obtained by propagating the pulse shown in the top row. 
    A level curve of $|E_x|$ at half its maximum value is shown as a blue line, to better visualize distortions of the energy density distribution.
    Blue (resp. red) dots mark the energy (resp. photon) centroids obtained from Eqs. \eqref{eq:REC_1st_order}, \eqref{eq:RPC_1st_order}. Cases presented are Gaussian pulses featuring \textbf{(a)} temporal chirp ($b=4$, $\gamma=2$), \textbf{(b)} wavefront rotation ($p=4$, $\gamma=2$), and \textbf{(c)} a STOV ($\ell=2$, $\gamma=1$) -- all cases use $1/(k_0 x_0) = 0.07$.
    }
    \label{fig:centroids}
\end{figure}
\\
We now give the total OAM, with respect to the origin. To first order, it reads
\begin{align}
\begin{split}
\vb{L} 
\approx 
\frac{1}{c}
\Bigg[
& \expval{
\vb{R}
}{\psi}
\cross \hat{\vb{z}}
+
\frac{\hat{\vb{z}}}{k_0}
\expval{
\vb{r}_\perp 
\cross
(-i\grad_\perp)
}{
\psi
}
\Bigg]
\end{split}
\label{eq:L_1st_order}
\end{align}
where $\vb{R} = (\vb{r}_\perp, -c\tau)$ denotes a space-time coordinate. 
The first term is transverse, and evidently of extrinsic nature as its value involves the arbitrary overall position of the pulse envelope. The second term is purely longitudinal and describes the OAM of regular, spatial vortex beams, generated by the well-known operator $\vb{r_\perp} \cross (-i\grad_\perp) = x(-i\partial_y) - y(-i\partial_x)$ \cite{nienhuis_1993_paraxial}. Equation \eqref{eq:L_1st_order} can be checked to reproduce the expression for $L_y$ found in \cite{porras_2024_clarification}. We remark that in the cited work, OAM was evaluated by integrating the \emph{total} angular momentum (AM) flux density, under the assumption that its spin part was negligible. Though expressions have been given for the orbital and spin AM flux densities separately, they are lengthy and explicitly involve vector potentials \cite{bliokh_2014_conservation}. Because orbital and spin parts naturally separate in reciprocal space \cite{cohen-tannoudji_1953_photons, li_2009_spin, mansuripur_2011_spin}, no such assumption is required in the present approach, and the spin angular momentum could equivalently be derived, relaxing our assumption of homogeneous linear polarization. In this more general case however, we note that additional polarization-dependent shifts of the centroids and contributions to $\vb{L}$ would appear \cite{supplementary}, owing to the spin Hall effect and spin-orbit interactions of light \cite{bliokh_2010_angular}, which are beyond the scope of the present work. 

\subsection{Shifts of the energy and photon centroids}

We turn to the energy and photon centroids. To first order, their positions at $t=0$ write
\begin{align}
    & \vb{r}_\text{EC}(t=0)
    \approx
    \expval{\vb{R}}
    +
    \frac{1}{k_0}
    \expval{
    -c\tau 
    \begin{pmatrix}
        -i\grad_\perp
        \\
        0
    \end{pmatrix}
    }
    \label{eq:REC_1st_order}
    \\
    \begin{split}
    & [\vb{r}_\text{PC}
    -
    \vb{r}_\text{EC}](t=0)
    \approx
    -\frac{1}{k_0}
    \Big\langle 
    \vb{R}
    \left[
    i\partial_\tau
    -
    \expval{i\partial_\tau} 
    \right]/c
    \Big\rangle
    \end{split}
    \label{eq:RPC_1st_order}
\end{align}
To our knowledge expressions \eqref{eq:REC_1st_order} and \eqref{eq:RPC_1st_order} are new; they conveniently provide $\vb{r}_\text{EC}$, a quantity usually defined as an average over space, from a pulse envelope prescribed at a single plane; and $\vb{r}_\text{PC}$, a quantity inherently defined in reciprocal space, from the pulse envelope provided in direct space. 

From \eqref{eq:REC_1st_order}, we observe that the transverse position of the energy centroid may differ from the apparent center $\expval{\vb{R}}$ of the pulse envelope in space-time: in the context of STOVs, it is linked to the subtle observation made by Porras \cite{porras_2024_comment,
porras_2024_clarification} that a STOV pulse symmetric in space-time $(x,t)$ is necessarily asymmetric when looked at in the spatial domain $(x,z)$. 
Meanwhile, \eqref{eq:RPC_1st_order} evidences an additional transverse shift between energy and photon centroids, illustrated by Bliokh with Bessel STOVs having circular symmetry in space \cite{bliokh_2023_orbital}. 
It originates from an asymmetry of the effective ``photon density'' within the pulse: a region where the local frequency is larger than the mean frequency is contributed by photons with higher energy.
In our first-order treatment, this inhomogeneity of the local frequency is captured by the density $-\Re{\psi^\star [i\partial_\tau - \expval{i\partial_\tau}]/\omega_0
 \,\psi}$, of which the barycenter generates the photon-centroid shift \eqref{eq:RPC_1st_order}.

In \autoref{tab:centroids}, we apply our formulae and compile the centroid shifts of certain Gaussian-type pulses -- for simplicity we take pulses structured in $(x,t)$ only and omit their $y$-dependence.
\begin{table}[b!]
\caption{Centroids of Gaussian-type pulses $\psi(x, t) = e^{-(x/x_0)^2 - (t/t_0)^2} \varphi(x, t)$ with space-time eccentricity $\gamma = ct_0/x_0$, featuring either temporal chirp, wavefront rotation (WR) or a spatiotemporal optical vortex (STOV). All pulses verify $\expval{\vb{R}} = 0$.} 
\centering
\makebox[\linewidth]{
\begin{tabular}{l c c c}
\hline\hline
\\[-1.5ex]
&
$\varphi(x, t)$
& $k_0\, \vb{r}_\text{EC}(t=0)$ 
& $k_0 \, \vb{r}_\text{PC}(t=0)$
\\[0.5ex]
\hline\hline
\\[-1ex]
Chirp & 
$e^{-ib(t/t_0)^2}$ 
& $\vb{0}$ 
& $b \, \hat{\vb{z}}$
\\[1.5ex]
WR &
$e^{-ip(x/x_0)(t/t_0)}$
& $\frac{1}{4} p \, \gamma \, \hat{\vb{x}}$ 
& 
$\frac{1}{4} p \, [\gamma - \gamma^{-1}] \, \hat{\vb{x}}$
\\[1.5ex]
STOV &
$\left[ \frac{t}{t_0}
- i \,\text{sgn}(\ell)
\frac{x}{x_0}
\right]^{\abs{\ell}}$
& $\frac{1}{2} \ell \, \gamma \, \hat{\vb{x}}$ 
& $\frac{1}{2} \ell \, [\gamma + \gamma^{-1}] \, \hat{\vb{x}}$
\\[1.5ex]
\hline
\end{tabular}}
\label{tab:centroids}
\end{table}
\begin{figure*}[t!]
    \centering
    \includegraphics[width=\linewidth, trim={0, 20pt, 0, 0}]{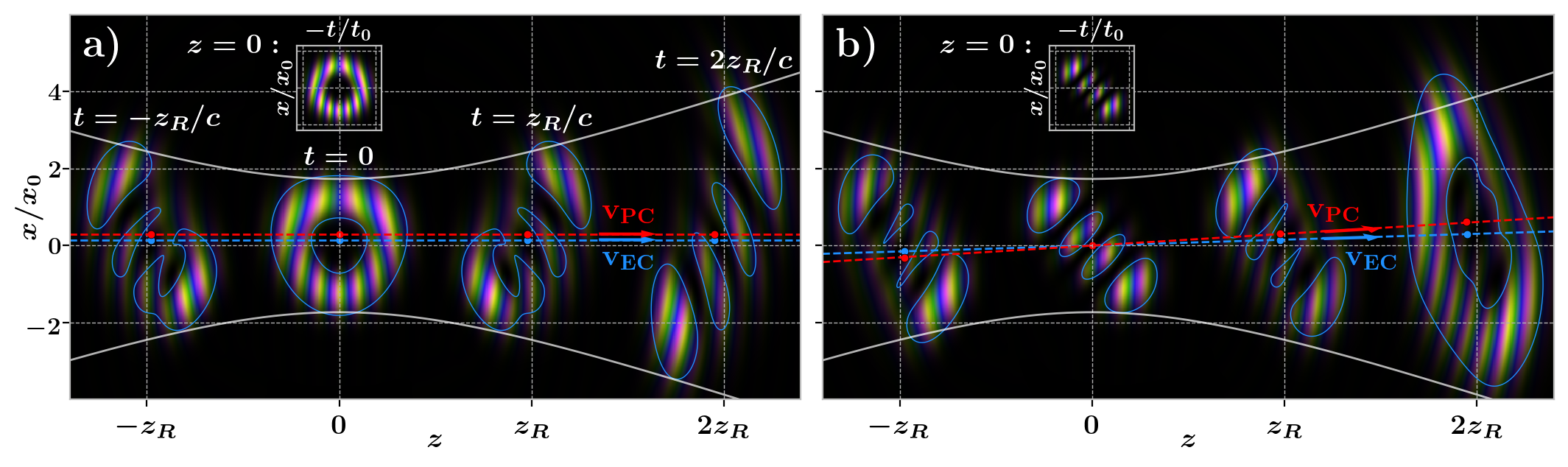}
    \caption{
    \textbf{Propagation of the energy and photon centroids.} \\
    Composite images of the complex electric field $E_x(x,z)$ at four instants of time, $t \in \{-z_R/c,\,0,\, z_R/c,\, 2z_R/c \}$, obtained by non-paraxial propagation of the space-time pulse $\psi(x,t) e^{-i\omega_0 t}$ realized at the plane $z=0$ (insets). The cases presented are \textbf{(a)} a symmetric STOV and \textbf{(b)} a THL mode, both with $\ell=3$, $\gamma = 1$ and $1/(k_0 x_0) = 0.1$.
    Blue (resp. red) dots mark the numerically evaluated energy (resp. photon) centroids at the four instants~; dotted lines indicate the trajectories of the centroids as obtained from the lowest-order formulae. White lines show the radius of a paraxial Laguerre-Gaussian beam $\sim x_0 \sqrt{1 + (z/z_R)^2} \sqrt{\ell}$ as a guide to the eye.
    }
    \label{fig:velocities}
\end{figure*}
These shifts are illustrated in \autoref{fig:centroids}, in which the top row shows the complex space-time pulses $\psi(x,t)e^{-i\omega_0 t}$ prescribed at $z=0$, while the bottom row shows the corresponding electric field component $E_x(x,z)$ in space at $t=0$, obtained by (non-paraxial) propagation of the prescribed pulse using Eqs. \eqref{eq:CAS} and \eqref{eq:PW_amplitude}.

Our first example is that of a chirped pulse with quadratic temporal phase, well-known as the result of passing a compressed pulse through dispersive optical elements. In \autoref{fig:centroids}(a), this results in lower frequencies ($\omega_-$) being located at the front of the pulse ($t<0$), and higher frequencies ($\omega_+$) at the rear ($t>0$). This produces an asymmetry in the photon density, inducing a relative longitudinal shift of the energy and photon centroids. For this pulse, which features no transverse phase gradient, the energy centroid coincides with the apparent center of the envelope, $\vb{r}_\text{EC} = \vb{R}$.

Our second example, in \autoref{fig:centroids}(b), is a pulse exhibiting wavefront rotation (WR). Such pulses have been used to drive the attosecond lighthouse effect in high-harmonic generation \cite{vincenti_2012_attosecond}, and were recently found to possess intrinsic transverse OAM \cite{porras_2024_procedure}.
Here, the local frequency varies along $x$: this again induces a shift of $\vb{r}_\text{PC}$ with respect to $\vb{r}_\text{EC}$, which gets displaced transversely, towards the lower frequencies. This time however, $\vb{r}_\text{EC}$ itself appears shifted with respect to the apparent center of the pulse: this is a manifestation of the second term in \eqref{eq:REC_1st_order}. It contributes because wavefronts at the front and rear of the pulse are tilted in opposite directions, such that the two ends of the pulse propagate at slight, opposite angles with respect to the $z$-axis. When looking at the field in space at $t=0$ (lower panel), the front has propagated slightly upwards to positive $z$, while the rear, propagating downwards, has not yet reached $z=0$. Overall, this deforms the energy distribution in space -- as highlighted by plotting one of its level curves (blue line) -- and results in a displacement of the energy centroid. 

These elementary principles can now be harnessed to review the centroid shifts in a STOV pulse, shown in \autoref{fig:centroids}(c). Here, the phase singularity at the center of the wavepacket induces a higher (resp. lower) local frequency at the bottom (resp. top). This explains the positive of shift of $\vb{r}_\text{PC}$ with respect to $\vb{r}_\text{EC}$. Meanwhile, the phase vortex also gives an opposite tilt to wavefronts at the front and rear of the pulse, resulting in a trigonal deformation of the energy density in space \cite{porras_2024_comment,
porras_2024_clarification}, and in the positive transverse shift of $\vb{r}_\text{EC}$. 

From definitions \eqref{eq:Li_EC} and \eqref{eq:Li_PC}, it follows that the value of intrinsic t-OAM is directly related to lateral shifts of the centroids. For the pulses of \autoref{tab:centroids} it can be checked that $\vb{P} \approx (U/c) \hat{\vb{z}}$ and that the total OAM with respect to the origin is $\vb{L} \approx \vb{0}$, hence their intrinsic OAM is directly dictated by the shifts given in the table as
\begin{align*}
    L_{i,\text{EC}}|_y
    \approx \frac{U}{c} x_\text{EC}
    \;\;\;\;\;
    L_{i,\text{PC}}|_y \approx \frac{U}{c} x_\text{PC}.
\end{align*}
For STOV pulses, the two distinct dependences of intrinsic OAM on the eccentricity $\gamma$ of the pulse have been much discussed already \cite{porras_2023_transverse, bliokh_2023_orbital}.
We remark that for the WR pulse, the photon-density and wavefront-tilt effects tend to shift the two centroids in opposite directions. Interestingly, for $\gamma < 1$ it is even possible for the two to be shifted on opposite sides of $\expval{\vb{R}}$, in which case the values of intrinsic OAM referenced to either centroid have opposite signs.

\subsection{Propagation of the centroids}

As captured by Eqs. \eqref{eq:rEC_vEC} and \eqref{eq:rPC_vPC}, the two centroids always propagate with constant velocities along rectilinear trajectories. Centroid velocities underpin conservation properties of intrinsic OAM in time: while $\vb{v}_\text{EC}$ is always aligned with $\vb{P}$, resulting in $\partial_t \vb{L}_{i,\text{EC}} = 0$, $\vb{v}_\text{PC}$ may not be, such that $\vb{L}_{i,\text{PC}}$ is not conserved in general. Although this fact has been emphasized by advocates of \eqref{eq:Li_EC} as a reason to discard this construction \cite{hancock_2024_spatiotemporal}, to our understanding it could still find relevance as a tool to describe certain OAM phenomena. In fact, because the space-time envelope of \emph{e.g.} a STOV wavepacket severely deforms upon propagation, it seems natural that its way of interacting with matter could change accordingly along the propagation axis: the interaction may then be more adequately quantified by a time-dependent parameter. Unlike the energy centroid, the photon centroid also has the property of being sensitive to frequency inhomogeneities within the pulse, which could prove important when dealing with dispersive materials.
\\

Limiting our calculations to first order in $\kappa$, the two centroids would appear to possess the same velocity $\vb{v}_\text{PC} \approx \vb{v}_\text{EC} = c^2 \vb{P}/U$, with $U$ and $\vb{P}$ given by \eqref{eq:U_1st_order} and \eqref{eq:P_1st_order}~; as already noted by Bliokh \cite{bliokh_2023_photon}, deviations between the two begin at order $\kappa^2$. 
Carrying our derivations to second order, we obtain
\begin{align}
    & \vb{v}_\text{EC}
    \approx
    c 
    \left[ 
    \hat{\vb{z}}
    +
    \frac{1}{k_0}
    \expval{
    \begin{pmatrix}
        -i\grad_\perp \\
        0 
    \end{pmatrix}
    }
    -
    \frac{1}{k_0^2}
    \expval{
    \begin{pmatrix}
        \grad_\perp \partial_\tau/c \\
        -\laplacian_\perp/2
    \end{pmatrix}
    }
    \right]
    \nonumber
    \\
    & 
    \vb{v}_\text{PC} - \vb{v}_\text{EC}
    \approx
-
c
\frac{1}{k_0^2}
    \expval{
    [i\partial_\tau - \expval{i\partial_\tau}]/c
    \begin{pmatrix}
         -i\grad_\perp \\
        0
    \end{pmatrix}
    }
    \label{eq:vPC_order2}
\end{align}
At this order, the difference between the two velocities may only be transverse -- the corresponding term measures the transverse propagation of all frequency components, weighted by their distance to the mean $\expval{i\partial_\tau}$ of the spectral envelope. Roughly speaking, if \emph{e.g.} higher frequencies are propagating in the positive $x$ direction more rapidly than lower frequencies do, the photon density (inversely proportional to $\omega$) will tend to accumulate at lower $x$, resulting in a negative contribution to the motion of the photon centroid along this axis. The time derivative of $\vb{L}_{i,\text{PC}}$ then reads
\begin{align*}
\partial_t
\vb{L}_{i,\text{PC}} 
& =
-\vb{v}_\text{PC} \cross \vb{P}
\\
& \approx
\frac{1}{k_0^2}
    \expval{
    [i\partial_\tau- \expval{i\partial_\tau}]/c 
    \begin{pmatrix}
         -i\grad_\perp \\
        0
    \end{pmatrix}
    }{\psi}\cross
    \hat{\vb{z}}.
\end{align*}
This lowest-order term affects only the transverse components of $\vb{L}_{i,\text{PC}}$. Owing to its second-order origin, appreciable relative variations of these components only accumulate after the pulse has propagated over a distance of order $\sim \kappa^{-2} k_0^{-1}$ (comparable to the Rayleigh length $z_R = k_0x_0^2/2$), far larger than its typical longitudinal size $\sim \kappa^{-1} k_0^{-1}$. Consequently, if $\vb{L}_{i,\text{PC}}$ is not a suitable quantity to describe propagation over long distances, its value could still be meaningful in the description of localized phenomena, occurring over sufficiently short longitudinal length scales.
\\

In \autoref{fig:velocities}, we illustrate the propagation of the two centroids in two cases relevant to the study of STOV t-OAM. For each, we show the electric field $E_x(x,z)$ at four instants of time, obtained by non-paraxial propagation of the pulse prescribed in space-time at $z=0$ (inset). At each instant, the numerical energy and photon centroids are shown as dots, while their approximate trajectories obtained from Eqs. \eqref{eq:REC_1st_order}, 
\eqref{eq:RPC_1st_order} and \eqref{eq:vPC_order2} are shown as dashed lines -- the corresponding analytical results are given in \autoref{tab:velocities}. 
\begin{table}[b!]
\caption{Centroid velocities for Gaussian-type pulses $\psi(x, t) = e^{-(x/x_0)^2 - (t/t_0)^2} \varphi(x, t)$ with space-time eccentricity $\gamma = ct_0/x_0$, realizing at focus either a spatiotemporal optical vortex (STOV) or a tilted Hermite-lobulated (THL) mode. $H_\ell\{\cdot\}$ is the Hermite polynomial of order $\ell$.} 
\centering
\makebox[\linewidth]{
\begin{tabular}{l c c c}
\hline\hline
\\[-1.5ex]
&
$\varphi(x, t)$
& $k_0^2\,[\vb{v}_\text{EC} - c\,\hat{\vb{z}}]$ 
& $k_0^2\,[\vb{v}_\text{PC}-\vb{v}_\text{EC} ]$
\\[0.5ex]
\hline\hline
\\[-1ex]
STOV &
$\left[ \frac{t}{t_0}
- i \,\text{sgn}(\ell)
\frac{x}{x_0}
\right]^{\abs{\ell}}$
& 
$- \frac{1 + \abs{\ell}}{2 x_0^2}
c\,\hat{\vb{z}}$ 
& 
$\vb{0}$
\\[3ex]
THL &
$
H_\abs{\ell} 
\left\{
\frac{t}{t_0}
+ \,\text{sgn}(\ell)
\frac{x}{x_0}
\right\}$
& 
\makecell{
$- \frac{1 + \abs{\ell}}{2 x_0^2} c\,\hat{\vb{z}}$
\\
$+ \frac{\ell}{x_0 ct_0}  c\,\hat{\vb{x}}$
}
& 
$\frac{\ell}{x_0 ct_0}  c\,\hat{\vb{x}}$
\\[3ex]
\hline
\end{tabular}}
\label{tab:velocities}
\end{table}
\\
In \autoref{fig:velocities}(a), the prescribed pulse is a symmetric STOV. In agreement with \autoref{tab:centroids}, the two centroids at $t=0$ are shifted differently along $x$. Their trajectories happen to be parallel however, such that for these specific pulses the intrinsic OAM with respect to either centroid are \emph{both} conserved in time. This was briefly remarked in \cite{bliokh_2025_ehrenfest}, and can be traced back to the fact that for these STOVs the plane-wave spectrum $|\widetilde{\vb{E}}(\vb{k})|^2$ is an even function of $k_x$ (see SM \cite{supplementary} for a short discussion).

A pulse for which the photon-centroid OAM would not be conserved was suggested in \cite{porras_2024_clarification}. In \autoref{fig:velocities}(b), we study a similar pulse, having a tilted, Hermite-lobulated (THL) profile in space-time. Such modes are of particular interest as the \emph{duals} to STOVs: they are the modes into which STOVs evolve under paraxial propagation to infinity, and vice-versa \cite{porras_2023_propagation, porras_2025_optics}. Here, although the two centroids coincide at
$t=0$ they propagate at different angles with respect to the $z$-axis. First, we note that $\vb{v}_\text{EC}$ has a second-order transverse component, related to a space-time coupling in the prescribed envelope $\sim\expval{\grad_\perp \partial_\tau/c}$. This post-paraxial term can be tentatively understood by noticing that at the plane $z=0$ the front of the pulse is lower than the rear: as the pulse traverses this plane, energy appears to be flowing slightly upwards overall, resulting in a second-order transverse contribution to $\vb{P}$, and thus to $\vb{v}_\text{EC}$ -- accordingly, this is not observed for the symmetric STOV of panel (a). Second, as the dual to a STOV pulse, this mode should deform into an elliptical STOV upon paraxial propagation to the far-field. In the non-paraxial propagation shown here, the envelope still evolves into a ring-shaped mode, with lower frequencies located at the top: this is associated with an additional transverse velocity of the photon centroid, such that the two trajectories are not parallel. As a result, only the energy-centroid intrinsic OAM is conserved, while the photon-centroid counterpart varies linearly with time.\\
Finally, we note that in both panels the longitudinal component of the two velocities is slightly less than $c$, due to the term $\sim \expval{\laplacian_\perp/2}$ in \eqref{eq:vPC_order2}. This is owed to the arc-like deformation of the pulse front that comes with post-paraxial diffraction -- this effect was already discussed in detail by Bliokh \cite{bliokh_2023_photon}.

\subsection{Intrinsic OAM per photon and optical vortices}

Recent literature on STOVs has emphasized the notion of intrinsic OAM \emph{per photon}. 
In our formalism, the intrinsic OAM per photon (in units of $\hbar$) with respect to either centroid read, to lowest order,
\begin{align}
\begin{split}
\vb{L}_{i,\text{EC}}^\text{ph} 
=
\frac{\vb{L}_{i,\text{EC}}}{N}
\approx &
\expval{
[\vb{R}-\expval{\vb{R}}] \cross
\begin{pmatrix}
         -i\grad_\perp \\
         0
\end{pmatrix}
}
\end{split}
\label{eq:LiEC_1st_order}
\\
\vb{L}_{i,\text{PC}}^\text{ph} 
=
\frac{\vb{L}_{i,\text{PC}}}{N}
\approx &
\expval{
[\vb{R}-\expval{\vb{R}}] \cross 
\begin{pmatrix}
         -i\grad_\perp \\
         i\partial_\tau/c
\end{pmatrix}
}.
\label{eq:LiPC_1st_order}
\end{align}
They are manifestly independent of the overall position $\expval{\vb{R}}$ of the prescribed pulse envelope, and therefore intrinsic. Both quantities write as the difference of one term of the form $\expval{\vb{R} \cross (-i\grad)}$ and the analogous term where the expectation values are taken separately, $\expval{\vb{R}} \cross \expval{-i\grad}$. 
\\
The energy-centroid construction only involves transverse momentum components $\grad_\perp$. Indeed, to first order the internal structure of a pulse will only contribute transverse components to the momentum density (see Eq. \eqref{eq:P_1st_order}) -- this is well illustrated by STOVs, for which it is known that the spatiotemporal phase vortex does not add additional momentum along the propagation direction \cite{hancock_2021_mode,
porras_2023_transverse}. In contrast, the photon-centroid construction features a complete (space-time) gradient $\grad_\text{st} = (\grad_\perp, -\partial_\tau/c)$. Yet, in obtaining this result we did not have to postulate the existence of an additional longitudinal momentum density \cite{hancock_2024_spatiotemporal}: inclusion of the temporal gradient is instead brought by the process of referencing to the photon centroid.
\begin{figure}[t!]
    \centering
    \includegraphics[width=\linewidth, trim={10pt, 20pt, 10pt, 0}]{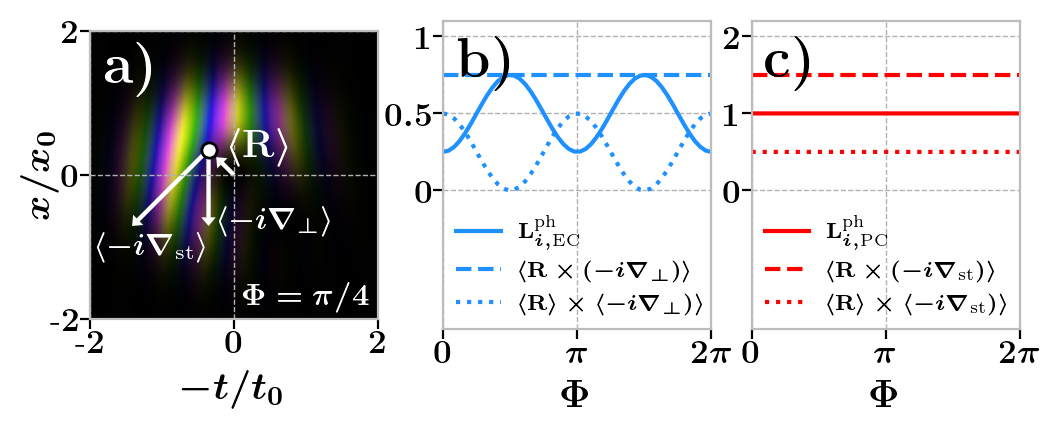}
    \caption{
    \textbf{Intrinsic OAM of a superposition of STOVs.} \\
    \textbf{(a)} Space-time mode corresponding to a superposition of two STOVs with topological charges $\ell_A=1$ and $\ell_B=2$, dephased by $\Phi=\pi/4$. A white dot and arrow marks the position of its apparent space-time center $\expval{\vb{R}}$. Other arrows indicate the relevant average phase gradients.
    Right: $y$-component of the intrinsic OAM per photon (full lines) for the same superposition of two STOVs as a function of their dephasing, referenced to the energy centroid \textbf{(b)} or to the photon centroid \textbf{(c)}. For each quantity, dashed and dotted lines indicate the two terms contributing to the intrinsic OAM per photon (see legend).
    }
    \label{fig:intrinsicOAM}
\end{figure}

The terminology \emph{per photon} may suggest that this is a property quantized at the level of single photons, in similar fashion to how the longitudinal OAM of regular vortex beams is quantized in integer steps \cite{molinaterriza_2007_twisted}. This tempting interpretation could be reinforced by recent studies on the upconversion of STOVs in harmonic or sum-frequency generation, which showed that their topological charge gets multiplied in accordance with the number of photons upconverted by the nonlinear process \cite{hancock_2021_second, gui_2021_second, fang_2021_controlling, gao_2023_spatiotemporal}. Yet, we can show with a simple argument that such intrinsic constructions are not quantities that photons in optical vortices carry in well-defined quanta. Consider a superposition of two STOV pulses, symmetric in space-time and with topological charges $\ell_A$ and $\ell_B$:
\begin{align*}
    \psi(x,t) & = \psi_{\ell_A}  + \psi_{\ell_B}  e^{i\Phi},
    \\
    \psi_\ell (x, t)
    & =
    \sqrt{\frac{2^\abs{\ell}}{\abs{\ell}!}}
    e^{-(x/x_0)^2-(ct/x_0)^2}
    \left[ \frac{ct}{x_0}
- i \,\text{sgn}(\ell)
\frac{x}{x_0}
\right]^{\abs{\ell}},
\end{align*}
where $\Phi$ is an arbitrary dephasing between the two pulses, and the normalization ensures both have the same energy $\braket{\psi_{\ell_A} } = \braket{\psi_{\ell_B} }$. For $\ell_A\neq\ell_B$ they are orthogonal, $\braket{\psi_{\ell_A} }{\psi_{\ell_B} } = 0$ and individually their intrinsic OAM per photon read
\begin{align*}
    L_{i,\text{EC}}^\text{ph}|_y(\psi_\ell) = \ell/2,
    \;\;\;\;\;
    L_{i,\text{PC}}^\text{ph}|_y(\psi_\ell) = \ell.
\end{align*}
If individual STOV photons all carried the same quantum of intrinsic OAM, a photon in the superposition should have the same probability of being detected with either of these two values, hence the average intrinsic OAM per photon of the superposition should be the average of that of each STOV taken separately. Yet, it can be shown that when $\abs{\ell_A - \ell_B} = 1$, this law does not hold. This is illustrated in \autoref{fig:intrinsicOAM} for $(\ell_A, \ell_B) = (1, 2)$. In \autoref{fig:intrinsicOAM}(a), we plot the space-time envelope of the superposition for the particular case $\Phi=\pi/4$. Although by symmetry each constituent STOV has its apparent center located at the origin $\expval{\vb{R}} = \vb{0}$, the same is not true for their superposition: this center gets shifted towards the region where they interfere constructively, which depends on their relative phase. In parallel, the superposition also has an average phase gradient, $\expval{-i \grad_\text{st}} \neq 0$. As a result, an extrinsic contribution to OAM is generated which must be subtracted to isolate the intrinsic part: terms of the form $\expval{\vb{R}} \cross \expval{-i \grad}$ in \eqref{eq:LiEC_1st_order} and \eqref{eq:LiPC_1st_order} contribute. In panels (b) and (c), we plot the corresponding intrinsic OAM per photon (full lines) as a function of $\Phi$, in the energy-centroid and photon-centroid frameworks, respectively. We immediately observe that their values match neither the naive $(\ell_A+\ell_B)/4 = 0.75$ nor $(\ell_A+\ell_B)/2 = 1.5$ in the respective cases. Each time this coincides with a non-vanishing term $\expval{\vb{R}} \cross \expval{-i \grad}$ (dotted lines). The effect is particularly striking in the energy-centroid picture, as keeping only the transverse phase gradient renders this term sensitive to the dephasing between the two STOVs. 

Most importantly, the same reasoning applies to regular, \emph{spatial} optical vortices (which could be obtained by replacing $-ct$ with $y$ in expressions above): there also, \emph{intrinsic} longitudinal OAM does not behave additively under a superposition. However, in this case the \emph{total} longitudinal OAM does: from Eq. \eqref{eq:L_1st_order}, we recall that it is generated by the operator $x(-i\partial_y) - y(-i\partial_x)$, for which vortex states are eigenmodes~; it then becomes trivial to show that the OAM per photon of the superposition is always the average of that for the two constituent vortex modes.

Our formalism hence highlights fundamental differences between transverse and longitudinal OAM of propagating pulses, and how they relate to optical vortices.
The total longitudinal OAM can be said to be carried in integer quanta by photons, as the field can be quantized on a basis of orthogonal eigenmodes of the corresponding operator with integer eigenvalues. These modes are spatial vortices, and for a given pure mode the OAM is naturally intrinsic, as $\vb{L} \propto \vb{P}$.
Meanwhile, as apparent in Eq. \eqref{eq:L_1st_order} the total transverse OAM of a pulse does not involve the analogous operator, and is also always origin-dependent -- the same observations were 
 made in theories of the t-OAM per unit length of monochromatic beam-like states of light \cite{aiello_2009_transverse, aiello_2010_transverse}. The various intrinsic constructions \eqref{eq:LiEC_1st_order}, \eqref{eq:LiPC_1st_order} that could be envisioned do not take the form of expectation values of a single operator for which vortex states would be the eigenmodes. As mentioned in \cite{bliokh_2025_ehrenfest}, such quantities do not appear to be observables measured by any physical process, and may be better considered as tools to gain intuitive insight into the OAM of optical fields. The ``quantization'' of STOV t-OAM mentioned in earlier studies \cite{hancock_2021_mode} stems merely from the fact that a basis of vortex states is itself indexed by a discrete parameter, the topological charge $\ell$ -- it appears meaningless from the standpoint of quantum optics and properties of single photons.
\\
Interestingly, the STOVs $\psi_\ell$ would be eigenmodes of $[\vb{R} \cross \expval{-i\grad_\text{st}}]|_y = -c\tau \partial_x - x \partial_\tau/c$, which is involved in the photon-centroid picture \eqref{eq:LiPC_1st_order} and matches the ``canonical'' OAM operator originally used by Bliokh \etal in early studies of STOVs \cite{bliokh_2012_spatiotemporal, bliokh_2021_spatiotemporal}. However, we emphasize that in the general case where $ \expval{\vb{R}} \cross \expval{-i\grad_\text{st}} \neq \vb{0}$, applying this operator alone to a pulse envelope neither generates its intrinsic nor its total t-OAM.

\section{Conclusions}
\label{sec-conclusion}

We have developed a rigorous theory of the energy and photon centroids of light pulses and the corresponding intrinsic orbital angular momenta, providing formulae conveniently applicable to an arbitrary pulse envelope. 
We have illustrated them on fields with simple structures to best showcase the principles dictating the trajectories of both centroids, but they could readily be applied to other pulses of recent interest, such as STOVs of higher radial orders \cite{liu_2024_spatiotemporala}, with arbitrary orientation of their singular lines \cite{wang_2021_engineering, zang_2022_spatiotemporal, porras_2023_control}, wavepackets threaded by several singularities \cite{wan_2022_photonic}
or optical hopfions \cite{wan_2022_scalar}, as well as experimentally measured pulses. Though we have focused here on near-paraxial fields, we note that exact, non-paraxial expressions for STOV pulses have recently been reported \cite{vo_2024_closed}, which may serve as a basis for post-paraxial investigations of t-OAM. We stress that the present theory is concerned with \emph{propagating} pulses, and does not describe the OAM that may come with \emph{static} transverse phase vortices, of the sort that can appear in simple superpositions of two monochromatic laser beams \cite{pas’ko_2001_transversal,
zhong_2024_toroidal} and of which the t-OAM has recently been detected mechanically \cite{hu_2023_structured} -- these structures are distinctly different from STOV pulses. 

We hope our formalism can help revealing fundamental differences between the total and intrinsic 
OAM of light pulses. Despite widespread use of the notion of intrinsic OAM \emph{per photon}, we observe that this quantity -- regardless of the centroid chosen to define it -- could not be carried in well-defined quanta by photons in optical vortices, be they spatial or spatiotemporal. Indeed, intrinsic OAM is not an observable associated to an operator for which optical vortices would be the eigenmodes. It should be noted that this conclusion is not a consequence of prescribing our pulse envelopes in space-time as $\psi(x,y,t)$: it could also be reached using a spatial formulation $\psi(x,y,z)$ as in \cite{porras_2024_clarification}. The situation contrasts sharply with the quantization of the total longitudinal OAM in spatial vortex beams, on which stands some of their celebrated applications \cite{erhard_2017_twisted}. This distinction may prove crucial for the proper interpretation of future experiments exploring the behaviour of STOVs in quantum optics, in the driving of electronic transitions \cite{babiker_2002_orbital, schmiegelow_2016_transfer} or in nonlinear optical processes, for which the upconversion of angular momenta quanta alongside photons frequently offers an invaluable framework \cite{roger_2013_non, zhdanova_2015_topological, kong_2017_controlling, luttmann_2023_nonlinear}.

\bibliography{allV2}

\end{document}